\documentclass[twocolumn]{aastex631}

\usepackage{comment}

\shorttitle{}

\shortauthors{Cabot et al.}
\begin{document}

\title{High-resolution Spectroscopic Reconnaissance of a Temperate Sub-Neptune}

\author[0000-0001-9749-6150]{Samuel H. C. Cabot}
\affiliation{Institute of Astronomy, University of Cambridge, Cambridge, UK}

\author[0000-0002-4869-000X]{Nikku Madhusudhan}
\affiliation{Institute of Astronomy, University of Cambridge, Cambridge, UK}

\email{Correspondence: nmadhu@ast.cam.ac.uk / shcc2@cam.ac.uk}

\author[0000-0001-6839-4569]{Savvas Constantinou}
\affiliation{Institute of Astronomy, University of Cambridge, Cambridge, UK}

\author[0000-0003-3993-4030]{Diana Valencia}
\affiliation{Department of Astronomy, University of Toronto, Canada}

\author[0000-0003-0489-1528]{Johanna M. Vos}
\affiliation{School of Physics, Trinity College Dublin, The University of Dublin, Dublin 2, Ireland}

\author[0000-0002-6939-0831]{Thomas Masseron}
\affiliation{Instituto de Astrofísica de Canarias, La Laguna, Tenerife, Spain}
\affiliation{Departamento de Astrofísica, Universidad de La Laguna, La Laguna, Tenerife, Spain}

\author[0000-0001-6839-4569]{Connor J. Cheverall}
\affiliation{Institute of Astronomy, University of Cambridge, Cambridge, UK}

\begin{abstract}
The study of temperate sub-Neptunes is the new frontier in exoplanetary science. A major development in the past year has been the first detection of carbon-bearing molecules in the atmosphere of a temperate sub-Neptune, K2-18~b, a possible Hycean world, with the James Webb Space Telescope (JWST). The JWST is poised to characterise the atmospheres of several other such planets with important implications for planetary processes in the temperate regime. Meanwhile, ground-based high-resolution spectroscopy has been highly successful in detecting chemical signatures of giant exoplanets, though low-mass planets have remained elusive. In the present work, we report the atmospheric reconnaissance of a temperate sub-Neptune using ground-based high-resolution transmission spectroscopy. The long orbital period and the low systemic velocity results in a low planetary radial velocity during transit, making this system a valuable testbed for high-resolution spectroscopy of temperate sub-Neptunes. We observe high-resolution time-series spectroscopy in the {\it H}- and {\it K}-bands during the planetary transit with the IGRINS instrument (R $\sim$45,000) on Gemini-South. Using observations from a single transit we find marginal evidence (2.2$\sigma$) for the presence of methane (CH$_4$) in the atmosphere and no evidence for ammonia (NH$_3$) despite its strong detectability for a cloud-free H$_2$-rich atmosphere. We assess our findings using injection tests with different atmospheric scenarios, and find them to be consistent with a high CH$_4$/NH$_3$ ratio and/or the presence of high-altitude clouds. Our results demonstrate the capability of Gemini-S/IGRINS for atmospheric characterization of temperate sub-Neptunes, and the complementarity between space- and ground-based facilities in this planetary regime. 
\end{abstract}


\section{Introduction} \label{sec:intro}
We are at the beginning of a new era in exoplanetary science. Exoplanet surveys have revealed that sub-Neptune planets with sizes between 1-4 R$_\oplus$ dominate the known exoplanet population \citep{fressin2013, Ricker2015, fulton2018}. In particular, a number of temperate (300-500 K) super-Earths and mini-Neptunes detected around bright M dwarfs are the most prized exoplanets from the TESS mission \citep[e.g.][]{Kostov2019, Gunther2019, cloutier2020, nowak2020,Demory2020}. The small stars lead to large transit depths and their brightness makes their planets readily conducive for atmospheric characterisation. These systems open up a rich regime of planetary processes in temperate conditions comparable to those in the solar system. Such high-value exoplanets form the ideal targets for atmospheric characterisation with JWST and large ground-based facilities.

Despite constituting the majority of the exoplanet population, sub-Neptune exoplanets have no analogue in the solar system. As such, they raise fundamental questions in planetary processes shaping their atmospheres, interiors, formation and habitability. The known masses and radii of sub-Neptune planets permit a wide range of internal structures, including the canonical super-Earths and mini-Neptunes, \citep{valencia2007, fortney2007, Seager2007, valencia2013}, as well as water worlds \citep{leger2004, Luque2021} and the recently-proposed Hycean worlds with deep H$_2$O oceans and H$_2$-rich atmospheres \citep{Madhusudhan2021}. Atmospheric characterisation of sub-Neptunes is necessary to resolve this degeneracy, as their atmospheric properties are expected to be influenced by the planetary interior compositions and surface-atmosphere interactions. For sub-Neptunes with H$_2$-dominated atmospheres, the presence and abundance of key trace molecules such as H$_2$O, CH$_4$, CO$_2$ and NH$_3$ can constrain whether or not the planet possesses a solid or liquid surface, or a Neptune-like deep atmosphere \citep{Yu2021, Hu2021, Tsai2021, Madhusudhan2023a}. This has recently been demonstrated with JWST observations of the candidate Hycean world K2-18~b, where the presence of CH$_4$ and CO$_2$, along with the non-detection of NH$_3$ were found to be consistent with the presence of an ocean surface \citep{Madhusudhan2023b}.

The study of temperate sub-Neptunes has entered a new era with the arrival of JWST. 
Until recently, atmospheric characterisation of temperate sub-Neptunes has been a formidable challenge. Observations with the Hubble Space Telescope (HST) WFC3 spectrograph (1.1-1.7 $\mu$m) led to nominal inferences of H$_2$-rich atmospheres with potential H$_2$O absorption in some temperate sub-Neptunes, e.g. K2-18~b \citep{Tsiaras2019, Benneke2019, Madhusudhan2020} and TOI-270~d \citep{MikalEvans2022}. However, it has been suggested that HST observations, due to their relatively limited spectral coverage and low resolution, may suffer from degeneracies between H$_2$O and CH$_4$ \citep{Blain2021, Bezard2022} and with the presence of starspots/faculae \citep{Barclay2021}. Recent observations with JWST have led to the first detections of prominent carbon-bearing molecules in the atmosphere of the temperate sub-Neptune K2-18~b, raising the prospects of it being a Hycean world as discussed above \citep{Madhusudhan2023b}. Similar JWST observations have been allocated for several other sub-Neptunes, which are well poised to revolutionise our understanding of their atmospheres and possible surface conditions.  

\begin{figure*}[t]
\noindent
\includegraphics[angle=0,width=\textwidth]{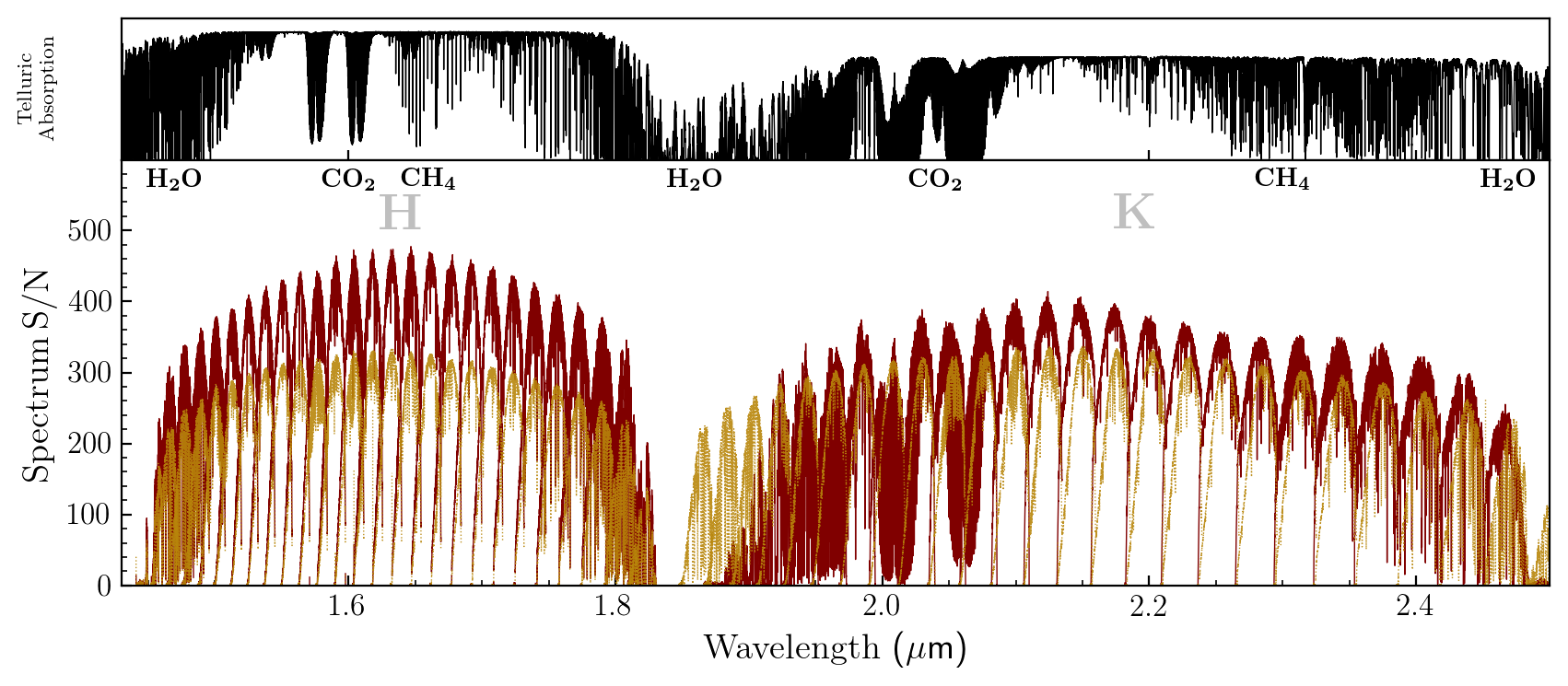}
\caption{Signal-to-noise (S/N) of the observations. The S/N of the first spectrum in our timeseries across the IGRINS wavelength range ($H$-band left, $K$-band right), obtained at airmass $z=1.08$, is shown in the maroon curves in the bottom panel. The S/N exceeds 400 in some parts of the stellar continuum. A fiducial telluric absorption spectrum is shown in the top panel \citep{Hinkle1995}. Some notable features are labeled by their corresponding molecular absorber. The lowest quality regions in the data primarily correspond to H$_2$O absorption zones, with CO$_2$ and CH$_4$ also having significant contributions. The repetitive arc-pattern is due to the blaze function instrumental response of individual spectral orders. For comparison, we plot the S/N of one spectrum used in the study of WASP-77~A~b \citep{Line2021}, marked by the dotted orange curve.
}
\label{fig:snr}
\end{figure*}

These developments make it timely to advance this frontier with high-resolution ground-based spectroscopy of sub-Neptune exoplanets. Firstly, of the numerous sub-Neptunes discovered by TESS, several orbit M dwarfs bright enough to be accessible to high-SNR high-resolution transit spectroscopy from the ground. Secondly, the recent chemical detections in K2-18b with JWST demonstrate both the H$_2$-rich nature of such atmospheres and the presence of prominent molecules, above any potential cloud decks, in abundances high enough to be detectable. Thirdly, ground-based high-resolution transit spectroscopy is now a proven technique for detecting chemical signatures in hot Jupiters \citep{Snellen2010, Birkby2013, Brogi2018, Sanchez-Lopez2019}, with recent efforts directed towards atmospheres of hot rocky exoplanets \citep{Jindal2020, Harper2023, Rasmussen2023}. Theoretical studies have demonstrated that the same is also feasible for mini-Neptunes orbiting bright M dwarfs \citep{gandhi2020, Hood2020}. Such ground-based observations have the potential to detect molecular features even in the presence of exoplanetary clouds that could otherwise obstruct the signatures in low-resolution space-based transmission spectra \citep{gandhi2020}. Finally, high-resolution transit spectroscopy offers an efficient way to study in depth both planet and parent star.  Thus, we are witnessing a new era where ground-based high-resolution spectroscopy can play a major role in characterising the atmospheres of low-mass temperate exoplanets, complementing observations with HST and JWST.

In the present work, we report atmospheric characterisation of a temperate sub-Neptune using ground-based high-resolution transit spectroscopy. The planet TOI-732~c or LTT~3780 c \citep{cloutier2020,nowak2020} has a mass of 8.04~M$_\mathrm{\Earth}$, a radius of 2.39~R$_\mathrm{\Earth}$, a zero-albedo equilibrium temperature of 359~K \citep{Bonfanti2023}, and orbits a nearby bright M dwarf (J=9.0, H=8.4) in a 12.3 day orbit. The bulk density of the planet allows for a degenerate set of internal structures, with possible atmospheric compositions including both a H$_2$-rich atmosphere and a H$_2$O-dominated steam atmosphere \citep{Bonfanti2023}. Moreover, given the low equilibrium temperature, TOI-732~c was also identified as a candidate Hycean world \citep{Madhusudhan2021}. The favourable system properties make the planet one of the most promising temperate sub-Neptunes for atmospheric characterisation with JWST \citep{Madhusudhan2021, Constantinou2022} and have been approved for upcoming JWST observations (GO Program 3557, PI: N. Madhusudhan). We conduct atmospheric reconnaissance of TOI-732~c using the IGRINS instrument on GEMINI-S, spanning a wavelength range 1.45-2.5~$\mu$m, which encompasses prominent high-resolution spectral features from the key molecules H$_2$O, CH$_4$ and NH$_3$.

Moreover, this system presents a challenging case study for HRCCS methods which have traditionally relied on the high velocity dispersion of close-in giant exoplanets \citep[e.g.][]{Snellen2010,Brogi2012,Sanchez-Lopez2019}. However, the HRCCS method is still effective in atmospheric characterisation of low-velocity planets (Cheverall \& Madhusudhan, in press), with our present target representing an extreme test case, given the low radial velocity during transit. The planet's long orbital period imparts a small Doppler shift in its atmospheric signature throughout transit, with the maximum radial velocity shift over the full transit of 1.9 km/s, i.e. sub-pixel velocity for IGRINS. Additionally, the systemic velocity of the system is consistent with zero, at 0.27 $\pm$ 0.34 km/s \citep{Bonfanti2023}.  Overall, the low net radial velocity of the planet during transit makes it particularly difficult to isolate the planetary atmospheric spectral features from the telluric and stellar spectral features. Therefore, our reconnaissance observations with such an end-member target provide first insights into the capability of current facilities for characterising temperate sub-Neptunes with HRCCS. 

In what follows, we present our GEMINI-S/IGRINS observations and the data analysis and modeling methodology in \S\ref{sec:obs}. We present the results of our HRCCS analysis in \S\ref{sec:res}, investigating the presence of key  molecular species in the atmosphere, as well as injection tests we carry out to assess the robustness of our findings. Finally, in section \S\ref{sec:disc}, we summarise our results and discuss our findings in the context of forthcoming observations of sub-Neptunes with JWST. 

\begin{figure*}
\noindent
\includegraphics[angle=0,width=\textwidth]{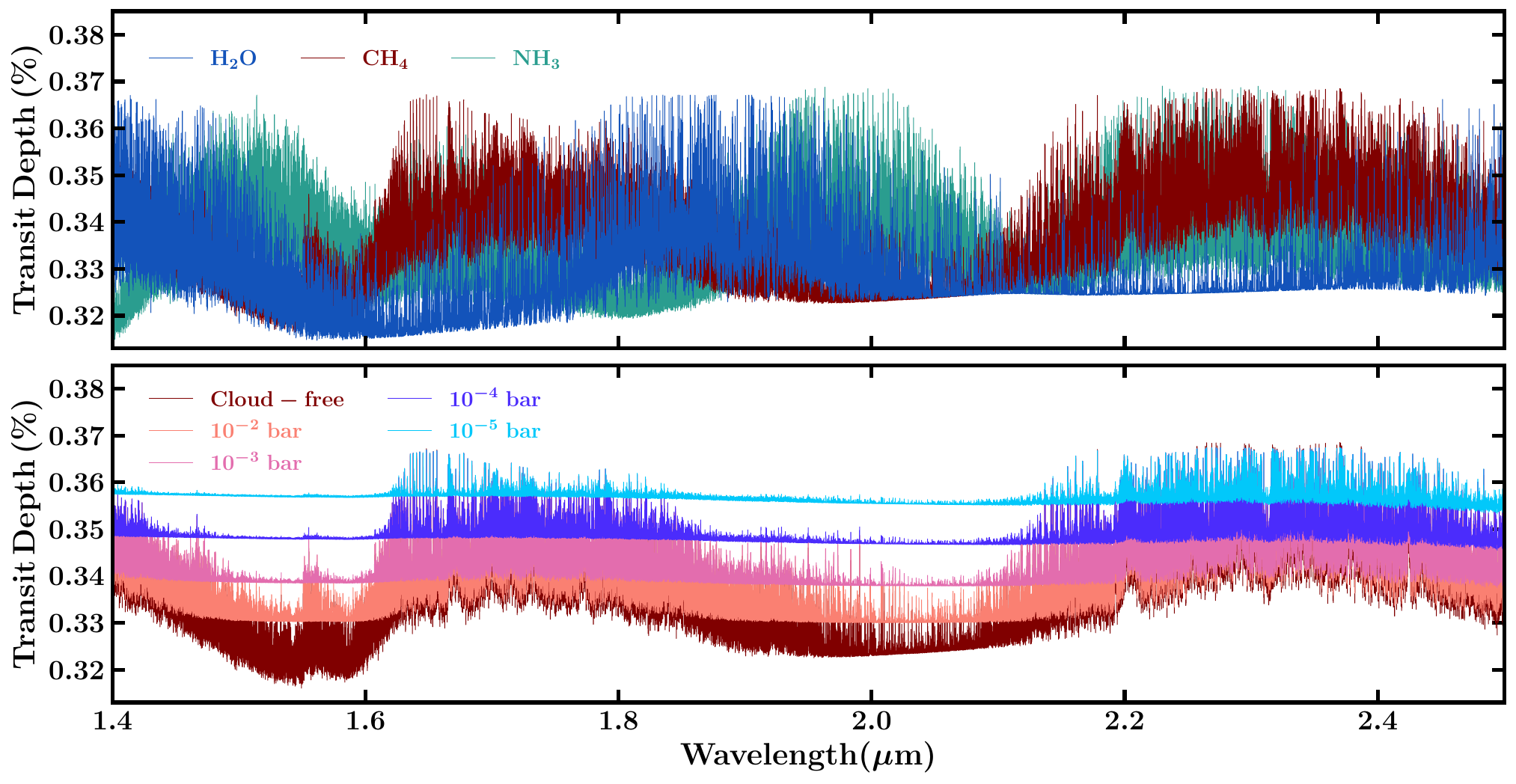}
\caption{Model transmission spectra used for cross-correlation. {\bf Top:} Transmission spectra with contributions from H$_2$O (blue), CH$_4$ (brown) and NH$_3$ (green), with the spectral baseline set by H$_2$-H$_2$ and H$_2$-He collision-induced absorption. {\bf Bottom:} Cloud-free and cloudy transmission spectra for the CH$_4$ model with a grey-opacity cloud deck placed at various cloud top pressures. The cloud-free spectrum shown in brown is the same in both figures.
}
\label{fig:spectra}
\end{figure*}

\section{Observations and Data Analysis} \label{sec:obs}

As follows, we report details of the IGRINS transmission spectra, including observing conditions and data quality diagnostics. We discuss pre-processing steps and methods in our HRCCS pipeline. Also provided is a description of our atmospheric models, which serve as templates in our cross-correlation analysis.

\subsection{Observations \& Reduction}

We observed time-series spectroscopy of the host star TOI-732 (LTT 3780) using the IGRINS infrared spectrograph on Gemini-South. The observations were carried out as part of the Gemini GO program GS-2021A-Q-201 (PI: Valencia), and included observations during two primary transits on the nights of 06 January 2021 (Night 1) and 24 February 2021 (Night 2), for $\sim$3 hours each. We assess data from both nights but report results only for Night 2 in the present work. The data from Night 1 is presented in the Appendix. The data from Night 2 displayed significantly higher quality, a result of both ideal photometric conditions as well as excellent airmass over the observing duration ($z$ spanned 1.08 to 1.18, with a minimum of 1.05). Our choice to exclude Night 1 is primarily due to its lower S/N and limited out-of-transit baseline (exposures from each night are compared in the Appendix). Our study aims to provide a proof of concept for HRCCS, which is best served by the higher S/N dataset. For completion, we performed our full analysis on Night 1, which did not yield any statistically significant detections, as discussed in the Appendix. 

{On Night 2}, we acquired in total 33 4-minute exposures (16 in-transit, 17 out-of-transit), where each exposure constitutes an A/B pair. IGRINS simultaneously captures $H$-band and $K$-band spectra, with coverage spanning $1.45-2.45$ $\mu$m at spectral resolution $R\sim 45,000$. Each spectral order was treated independently throughout our analysis until aggregation of their respective cross-correlation functions (CCFs, see \S\ref{subsec:ccf}). IGRINS has 54 spectral orders in total. Although, only a subset of the orders outside of severe telluric contaminated zones were used (see below for masking details). 

The data reduction, spectral extraction, and wavelength calibration were done using the IGRINS Pipeline Package (PLP) \citep{Lee2017} by the IGRINS instrument team. The resulting spectral datacubes were used for the HRCCS analysis in the present work. A telluric standard A0V star was observed before and after the science exposures. The time-series spectra were divided by the spectrum of the later telluric standard which provides a preliminary continuum normalization (these data correspond to the $\texttt{spec\_a0v}$ reduced PLP products). The typical pixel S/N was $\sim 300$ (Figure~\ref{fig:snr}). However, the S/N exceeded $400$ in the stellar continuum, near the peak of the instrument response and outside of telluric bands.

\subsection{Cleaning \& Normalisation}

The most evident features in the source spectra are bands of strong telluric contamination, predominantly due to absorption by H$_2$O, CH$_4$, and CO$_2$, as well as the blaze function inherent to the instrument's spectral response. The telluric standard division essentially removes the blaze component, and significantly reduces the presence of tellurics. The telluric standard used for the initial division was the one observed after science exposures in the standard IGRINS observing procedure.  Additional pre-processing steps were taken in order to 
remove remaining broadband variations, time-dependent features, as well as the most contaminated wavelength ranges. We place the spectra into a 2D array, each column representing a wavelength bin and each row representing an exposure (or, equivalently, time or orbital phase). First, we divide each spectrum by its median value in order to normalize throughput. Although corrected spectra were already continuum-normalized, we continued to remove any remaining broadband trends. We apply a sliding 31-pixel filter which selects for the 95$^{\rm th}$-percentile value within its window. These values serve as approximate samples of the spectrum's continuum. We then fit a quadratic function to the these values, and finally divide the spectrum by this fit. 

We proceed to mask orders that fell in significant water absorption bands, in order to mitigate the effects of telluric contamination. Specifically, the following orders were masked: 1, 2, 3, 4, 5, 6, 25, 26, 27, 28, 29, 30, 31, 32, 33, 34, 35, 36, 37, 38, 39, 49, 50, 51, 52, 53, 54. More columns were masked in the remaining orders based on their contamination level. The masking consistently included the first 400 pixels and the last 100 pixels, which were affected by edge effects. We used a blaze-corrected, telluric standard spectrum taken on Night 1 to mask additional wavelength channels. Specifically, any channel with flux below 0.99 in the blaze-corrected spectrum was flagged and subsequently masked from our primary science spectra. In addition to remaining, shallow, telluric features, the M-dwarf host imparts a rich molecular absorption spectrum. These lines are removed by detrending the timeseries spectra as follows.

\subsection{Detrending}
\label{subsec:detrending}

Each spectral order bears a unique distribution of exoplanetary, stellar, and telluric features. We detrend the spectra in order to isolate the exoplanet's atmospheric signal. 
For close-in HRCCS targets, the atmospheric signal is significantly Doppler shifted throughout the transit. In contrast, stellar and telluric features are approximately stationary, and they are present in all frames. They may be efficiently removed via Singular Value Decomposition (SVD) \citep{deKok2013}. SVD can also identify and remove telluric trends related to the time-variable column density of the absorber. In our case, the planetary Doppler shift is a moderate $\sim 0.12$ km s$^{-1}$ (0.06 pixels) between consecutive exposures. However, the atmospheric signal is present during in-transit exposures only. Our out-of-transit baseline therefore establishes telluric and stellar features as the dominant trends, and the atmospheric signal contributes negligibly to the leading order singular vectors (Appendix \ref{sec:oot}). In our implementation, we compute a low-rank approximation of the spectra with SVD, divide the data by this approximation, and finally subtract unity, thereby isolating features in the transmission spectrum. Optimizations to this procedure are discussed in \S\ref{subsec:ccf}. 

It is important to note that the original spectra are in the geocentric rest frame, and as a result, stellar features are further Doppler shifted by the barycentric correction velocity. This quantity varies by $\sim 0.34$ km s$^{-1}$ throughout the 33-frame timeseries. In-transit spectra experience $\sim 0.16$ km s$^{-1}$ of this drift, which also shifts the atmospheric signal. Nevertheless, we find that the stellar features are largely removed in the detrending process, and our injection/recovery procedure demonstrates our robustness to these effects.

\subsection{Model Spectra for Cross-correlation}
\label{subsec:model_spectra}

We generate a library of simulated transmission spectra for cross correlation with the data (Figure~\ref{fig:spectra}) using the AURA atmospheric modelling and retrieval framework \citep{Pinhas2018, Welbanks2019, Constantinou2023}. The atmosphere at the day-night terminator is modelled in plane-parallel geometry under the assumption of hydrostatic equilibrium and isothermal temperature structure. Transmission spectra are generated through a radiative transfer calculation at a resolution of R = 300,000.

The chemical composition of the model atmosphere is uniform and set through the individual mixing ratios of trace molecular species, with the remainder made up of H$_2$ and He in solar elemental ratio. The spectral baseline of the resulting transmission spectrum is provided by H$_2$-H$_2$ and H$_2$-He collision-induced absorption \citep{borysow1988,orton2007,abel2011,richard2012}. The model can also include spectral absorption contributions from H$_2$O \citep{barber2006, rothman2010}, CH$_4$ \citep{yurchenko2014} and NH$_3$ \citep{yurchenko2011}. All three molecules are expected to be primary elemental carriers in temperate H$_2$-dominated atmospheres \citep{burrows1999, Lodders2002}. The cross-sections of all molecular species are computed following \citet{Gandhi2017} and \citet{gandhi2020} and account for pressure broadening due to H$_2$. We specifically consider mixing ratios corresponding to 10$\times$ solar elemental abundances under thermochemical equilibrium. These are 10$^{-2}$ for H$_2$O, $5 \times 10^{-3}$ for CH$_4$ and 10$^{-3}$ for NH$_3$. The isothermal temperature in the atmosphere is nominally assumed to be 350 K, consistent with the zero-albedo equilibrium temperature of the planet with efficient day-night redistribution. 

In addition to spectral contributions from gaseous species, the model atmosphere can also include extinction arising from atmospheric aerosols. We follow a parametric approach, modelling atmospheric aerosols as a grey opacity across the observed wavelength range, described by a cloud top pressure $P_\mathrm{c}$. For all models, we additionally include extinction arising due to H$_2$ Rayleigh scattering.

\subsection{CCF Optimization}
\label{subsec:ccf}

\begin{figure*}[t]
\noindent
\includegraphics[angle=0,width=\textwidth]{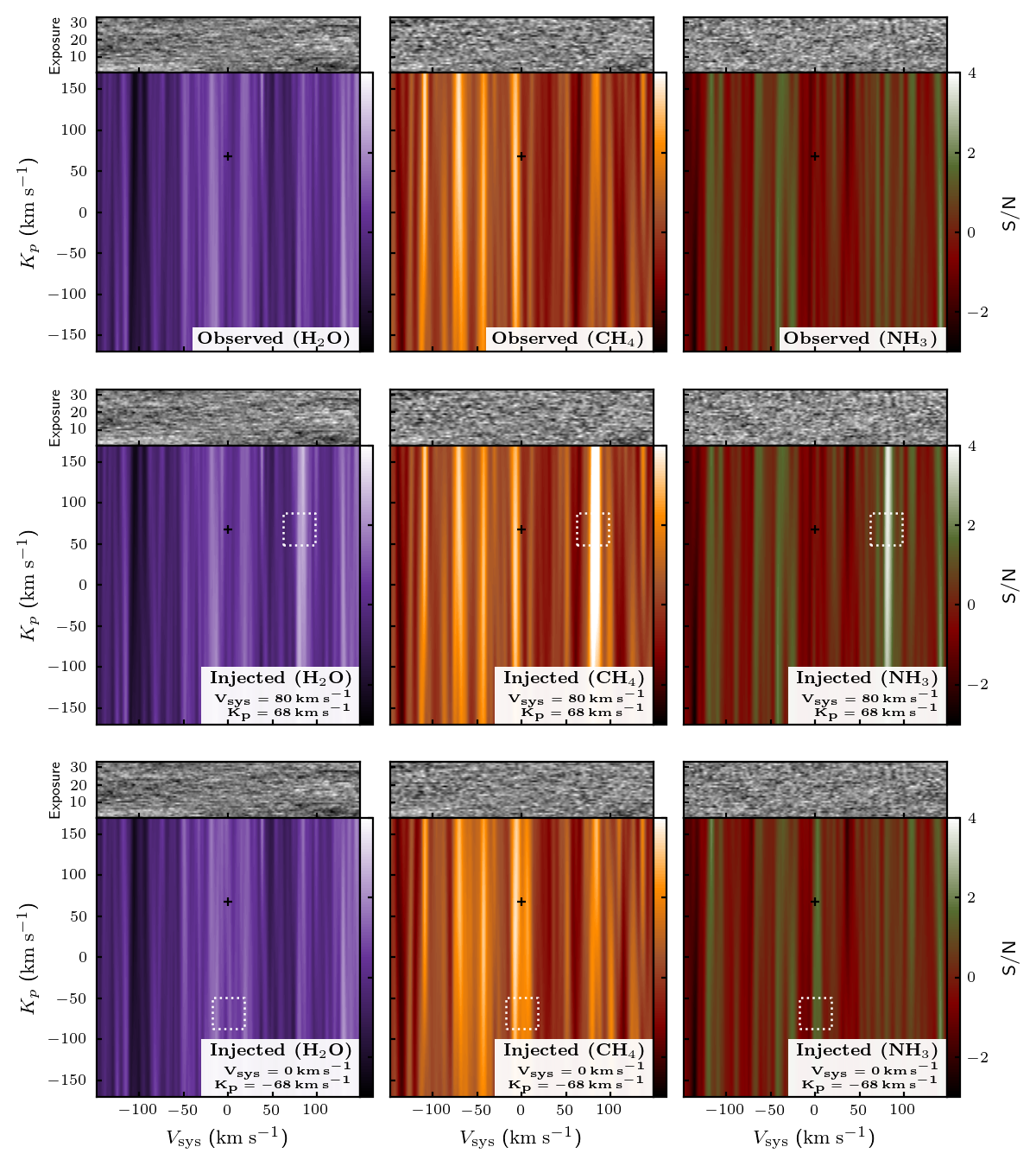}
\caption{{\bf Top}: Molecular inferences from cross-correlating the detrended spectra with our model templates. Results are shown for three species (CH$_4$, NH$_3$, and H$_2$O). Bright regions in these maps correspond to peaks in the CCF. For species in the atmosphere of the exoplanet, a positive signal is expected at the true orbital parameters, i.e. $K_{\rm p} \approx 68$ km s$^{-1}$ and $V_{\rm sys} \approx 0$ km s$^{-1}$ \citep[][indicated by the black ``+"]{nowak2020, Bonfanti2023}. The standard deviation across other velocity combinations serves as a proxy for noise when deriving the S/N of the signal. Note, each panel spans the same S/N color range (-3 to +4). The gray-scale arrays at the top of each panel depict the CCFs for each exposure. These CCFs are Doppler shifted by the planet's velocity under an assumed ($K_{\rm p}$, $V_{\rm sys}$)-pair, and subsequently stacked, which produces the color-scale S/N maps. {\bf Middle}: Same as the top row, except an artificial transmission spectrum was multiplied into the spectra at the start of data analysis. The injection assumes an orbit specified by $K_{\rm p} = 68$ km s$^{-1}$ and $V_{\rm sys} = 80$ km s$^{-1}$, as indicated by the dotted white boxes in each panel. {\bf Bottom}: Same as middle row, except the injection is specified by $K_{\rm p} = -68$ km s$^{-1}$ and $V_{\rm sys} = 0$ km s$^{-1}$ so as to more closely mimic the statistical properties of the observed signal.
}
\label{fig:maps}
\end{figure*}

Following standard HRCCS practices \citep[e.g.,][]{Brogi2012}, we cross-correlate our atmospheric models with the detrended transmission spectra. This step accumulates signal from the multitude of weak atmospheric features. Our analysis uses the X-COR pipeline \citep{Hawker2018, Cabot2019}, with details as follows. The cross-correlation function (CCF) between the Doppler-shifted model template $m(\lambda|v)$ and observed transmission spectra $f(\lambda, t)$ is given as:

\begin{equation}
    {\rm CCF}(v, t)=
    \frac{\sum_\lambda f(\lambda, t)w(\lambda)}
    {\sum_\lambda w(\lambda)} \, \, , \, \, w(\lambda) = \frac{m(\lambda| v)}{\sigma^2(\lambda)} 
    \label{eq:cc}
\end{equation}
where $\lambda$ represents the wavelength channel, $t$ the time of exposure, and $v$ the planetary radial velocity. The sum is weighted by a factor $w$ which includes the inverse of each channel's time-axis (column) variance $\sigma^2(\lambda)$, thereby minimizing contributions from cores of stellar or telluric lines. From each row of the CCF (i.e., corresponding to each exposure), we further subtract its median value (the result at this stage is shown in the gray-scale arrays in Figure~\ref{fig:maps}). Finally, we shift each row of the CCF by the expected planetary velocity, given by $v(t) = K_{p} \sin \phi + V_{\rm sys} + v_{\rm bary}$ for semi-amplitude $K_p$, systemic velocity $V_{\rm sys}$, and barycentric correction velocity $v_{\rm bary}$. We then sum the CCFs across in-transit exposures. This process was repeated for other orbital solutions (i.e., alternate values for $K_p$ and $V_{\rm sys}$), which provide a noise distribution for gauging significance.

Detrending parameters can significantly influence the statistical significance, and robustness, of observed signals \citep{Cabot2019, Cheverall2023}. In our analysis, we determine the number of singular vectors ($n_{\rm SV}$) to remove for each order. We choose $n_{\rm SV}$ based on $\Delta$CCF optimization \citep{Spring2022, Holmberg2022, Cheverall2023}: For each order, we select $n_{\rm SV}$ which maximizes the S/N of $\Delta$CCF. Here, $\Delta {\rm CCF} = {\rm CCF_{inj}} - {\rm CCF_{obs}}$, where ${\rm CCF_{obs}}$ is the CCF computed from the observed data, and ${\rm CCF_{inj}}$ is the same plus an additional transmission model injected at 1$\times$ nominal strength under the system's known $K_{\rm p} \approx 68$ km s$^{-1}$ and $V_{\rm sys} \approx 0$ km s$^{-1}$ \citep{nowak2020, Bonfanti2023}. We find better injection recovery by optimizing the peak of the $\Delta {\rm CCF}$ as opposed to the value at the exact injected ($K_{\rm p}$, $V_{\rm sys}$). We adopt this practice throughout our optimizations. As is standard, the noise baseline is derived from ${\rm CCF_{obs}}$ (computed row-wise for each frame, and excluding the region $|V_{\rm sys}| < 100$ km s$^{-1}$). This approach maximizes recovery of the injected signal while mitigating coherent accumulation of noise from different spectral orders. Note, we repeat the full optimization and cross-correlation procedure for each atmospheric model considered.

\section{Results}
\label{sec:res}

In this section we present our findings focused primarily on a search for key molecular species in the atmosphere of TOI-732~c and to assess the robustness of the same. We first conduct HRCCS analyses in search of H$_2$O, CH$_4$ and NH$_3$, which are expected to be the primary carbon-, oxygen- and nitrogen-carrying molecules in temperate H$_2$-rich atmospheres. CO$_2$ models were also investigated; however, we did not recover a significant detection. Our analysis is less sensitive to CO$_2$ since the planetary spectral features are fewer in number compared to other molecules, and they overlap with deep telluric CO$_2$ lines. Most CO$_2$ features are removed by our mask.

To assess the robustness of our results, we also conduct a series of injection tests, whereby we multiply a model transmission spectrum into the original data and recover the artificial signal with cross-correlation. The results of this analysis let us gauge the sensitivity of our observations, and assess the plausibility of our results. We extend the injection tests to involve cloudy atmospheric models, in order to explore physical conditions that are consistent with our cross-correlation results.

\subsection{Data Quality}
\label{subsec:quality}

At the outset, we highlight the quality of our observations against those obtained in previous studies. TOI-732 is representative of the bright, sub-Neptune hosting M dwarfs discovered by TESS. However, at $H=8.44$, it is somewhat fainter compared to the typical HRCCS targets hosting giant exoplanets. For example, the well studied hot Jupiter hosts HD 189733 and HD 209458, which were subjects of NIR atmospheric transmission spectroscopy \citep[e.g.]{Snellen2010, Brogi2018, Sanchez-Lopez2019}, have $H$-band magnitudes 5.59 and 6.37, respectively \citep{Cutri2003}. Nevertheless, our spectra attained remarkably high S/N stemming from favorable observing conditions. In particular, observations were conduced at low airmass; mid-transit corresponded to $z = 1.068$. In Figure~\ref{fig:snr}, we plot the S/N of one representative spectrum from our study (red curve), which reached over 400 in some of the orders. This quality is comparable to that obtained using the same instrument for a confident detection of multiple molecules in a hot Jupiter orbiting the Sun-like star WASP-77~A ($H=8.37$) using emission spectroscopy \citep{Line2021}. The S/N for one representative spectrum of WASP-77~A is also plotted in Figure~\ref{fig:snr}. We do note there are some differences between our analysis and \citet{Line2021}. The latter probes a much stronger signal originating from dayside emission features, and the atmospheric features exhibit significantly greater Doppler shift than in the present case. Nevertheless, and as we will show in \S\ref{subsec:injection}, the high fidelity of our data enable our sensitivity to molecular species in the atmosphere of TOI-732~c.

\begin{figure}[t]
\noindent
\includegraphics[angle=0,width=\linewidth]{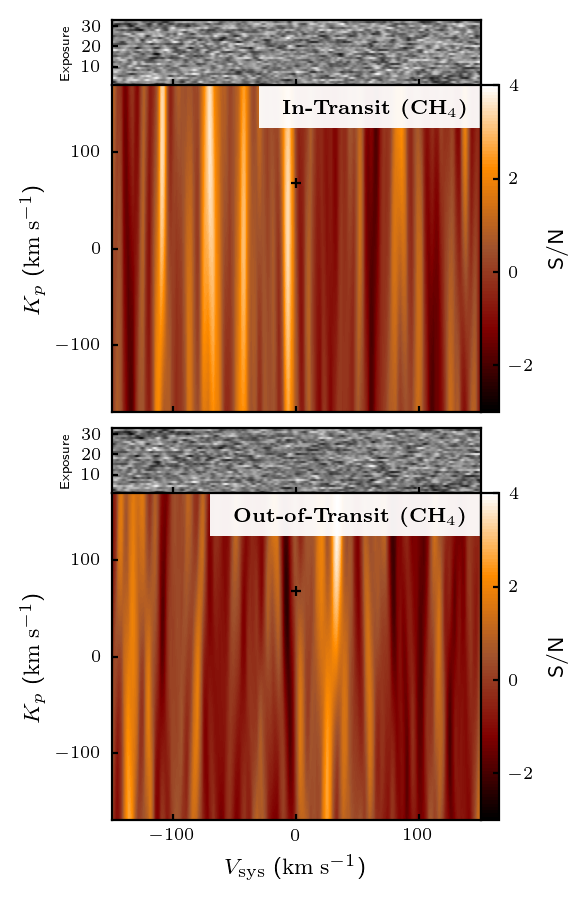}
\caption{{\bf Top}: S/N map showing a marginal signal for CH$_4$, identical to the corresponding panel in Figure~\ref{fig:maps}. {\bf Bottom}: Same as top, except that out-of-transit frames were used throughout $\Delta$CCF optimization and CCF stacking, instead of in-transit frames. There is no positive enhancement in the resulting S/N map near the expected planet location, lending evidence that the nominal signal is not present in the out-of-transit frames.
}
\label{fig:flip}
\end{figure}

\subsection{Molecular Inferences}
\label{subsec:molecular_inferences}

We perform cross-correlation with atmospheric models of CH$_4$, NH$_3$, and H$_2$O --- each a key molecule expected in temperate H$_2$-rich atmospheres. We do not detect any of the molecules at high statistical significance, but see nominal evidence for some trends. We report marginal (2-3$\sigma$) evidence for CH$_4$, based on CCF enhancement near the expected $K_{\rm p}$ and $V_{\rm sys}$ (the values of $n_{\rm SV}$ for each order, based on $\Delta$CCF optimization, are listed in Table~\ref{tab:svd}). The enhancement matches the morphology of signals recovered in injection tests (in particular, the cloudy models explored in \S\ref{subsec:injection_cloudy}). We find no evidence of either NH$_3$ or H$_2$O. In the case of NH$_3$, injection tests demonstrate high sensitivity to absorption by the molecule. However, our masking has removed regions with the strongest H$_2$O absoprtion features, and the remaining coverage has low sensitivity to the molecule. The top row of Figure~\ref{fig:maps} shows S/N maps from the cross-correlation for pairs of ($K_{\rm p}$, $V_{\rm sys}$). 

\begin{figure}[t]
\noindent
\centering
\includegraphics[angle=0,width=\linewidth]{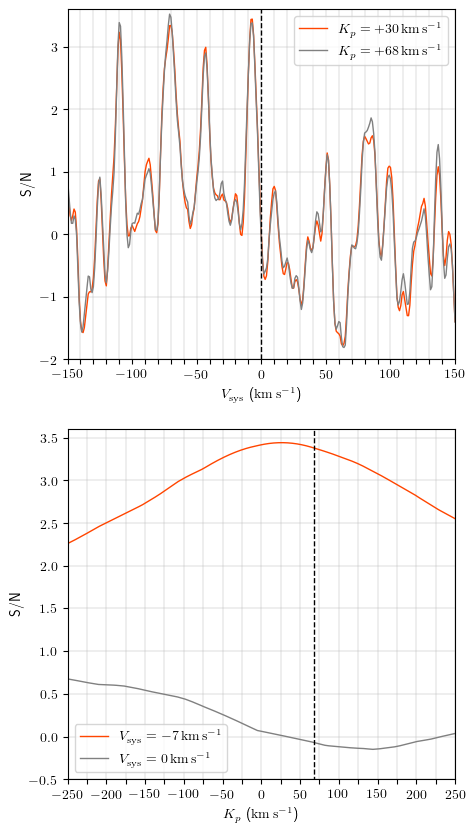}
\caption{The top and bottom panels show 1D slices of S/N near the tentative CH$_4$ signal. The orange curves correspond to the peak in the $K_{\rm p}$--$V_{\rm sys}$ plane at $K_{\rm p} = +30$ km s$^{-1}$ and $V_{\rm sys} = -7$ km s$^{-1}$, respectively. For reference, the gray curves correspond to the physical orbital parameters of TOI-732~b, $K_{\rm p} = +68$ km s$^{-1}$ and $V_{\rm sys} = 0$ km s$^{-1}$, respectively (also denoted by vertical dashed lines), indicating the velocity offset of the putative peak signal. As shown in the top panel, we find multiple peaks of comparable strength to our putative signal, albeit at much larger velocity offsets. These additional peaks reduce the statistical confidence of the CH$_4$ signal.}
\label{fig:1d}
\end{figure}

The weak CH$_4$ signal manifests as a broad, vertical enhancement in its corresponding S/N map, near the planet's $K_{\rm p}$ and $V_{\rm sys}$. It is slightly offset from the expected $V_{\rm sys}$, with significance reaching $2.2\sigma$ to $3.4\sigma$ at offsets of $-4$ to $-7$ km s$^{-1}$ (we show 1D profiles at the the signal's peak in Figure~\ref{fig:1d}). As such, we conservatively quote evidence for the molecule at the $2.2\sigma$ level. This confidence is supported by a bootstrap-based robustness test, included in Appendix~\ref{subsec:bootstrapped}. Vertical structure in the signal is expected, and is also present for injected signals described in \S\ref{subsec:injection} and \S\ref{subsec:injection_cloudy}. This structure arises because the transit comprises a very small portion of the planet's orbit ($-0.0023 \lesssim \phi \lesssim +0.0023$), and the change in the planet's radial velocity  $\Delta v_p$ is only 1.84 km s$^{-1}$. The process of shifting CCFs by the planet's velocity, and then aggregating across exposures, provides quantitatively similar results for various $K_{\rm p}$. In other words, $K_{\rm p}$ is poorly constrained from the atmospheric cross-correlation analysis, even though it is known at high-precision based on the system parameters derived from radial velocities and transit photometry \citep{nowak2020, cloutier2020, Bonfanti2023}. As seen in Figure~\ref{fig:1d}, at the peak $V_{\rm sys} = -7$ km s$^{-1}$, values of $K_{\rm p}$ ranging from -213 km s$^{-1}$ up to the top boundary of the velocity $K_{\rm p}$ space exhibit S/N in excess of $2.4\sigma$ (i.e., within $1\sigma$ of the peak). On the other hand, $V_{\rm sys}$ shows a tighter distribution constrained from -10 km s$^{-1}$ to -5 km s$^{-1}$, again localized around the peak.

\begin{figure*}[t]
\noindent
\includegraphics[angle=0,width=\textwidth]{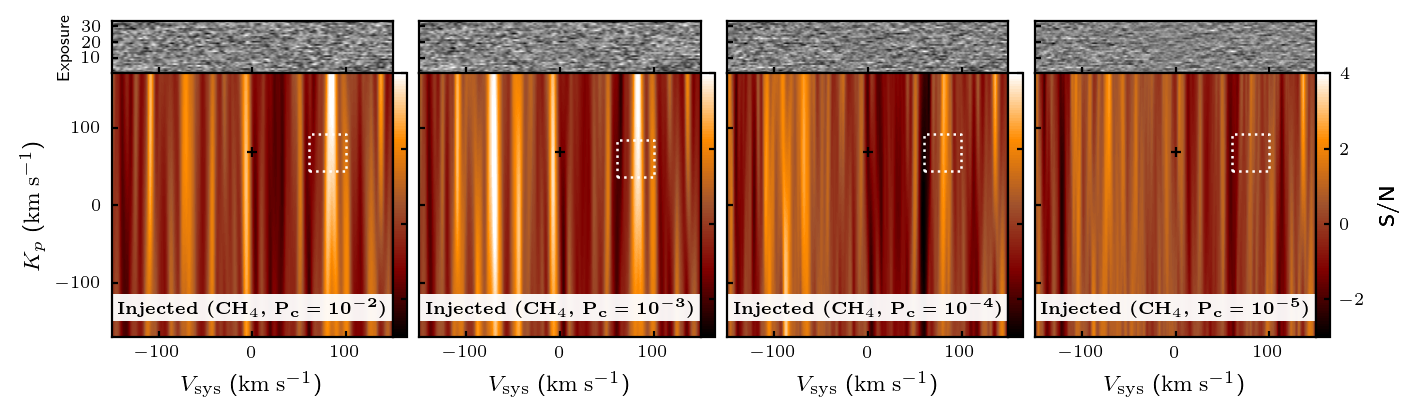}
\caption{S/N maps for injection/recovery tests for cloudy atmospheric models. These maps are generated in a manner similar to those shown in Figure~\ref{fig:maps}. Model transmission spectra were injected into the data at orbital parameters $K_{\rm p} = 68$ km s$^{-1}$ and $V_{\rm sys} = 80$ km s$^{-1}$, indicated by the dotted white box. The only molecule considered is CH$_4$, with a varying cloud-top pressure of 10$^{-2}$, 10$^{-3}$, 10$^{-4}$, and 10$^{-5}$ bar, in the panels proceeding from left to right. Increasing the altitude of the cloud deck, i.e. decreasing the cloud-top pressure, reduces the amplitudes of the spectral features and yields consistently weaker recovered signals.}
\label{fig:maps_cloudy}
\end{figure*}

We perform an additional robustness test on the tentative CH$_4$ signal by repeating the analysis described above, but swapping the in-transit frames with out-of-transit frames. Specifically, in $\Delta$CCF optimization, the artificial signal is injected only into out-of-transit frames, and only the CCFs of out-of-transit frames are combined when we aggregate the planetary signal. All 17 out-of-transit frames are used in this test. The result is shown in Figure~\ref{fig:flip}, and indicates that the CH$_4$ signal is not present in the out-of-transit spectra. This finding lends evidence that the nominal signal is not telluric in origin (or stellar, perhaps through a spurious peak in the cross-correlation of stellar lines and CH$_4$ features). 

The slight offset in the CH$_4$ signal is not immediately reconcilable, but could be explained by a number of factors. Similar blueshifted offsets on the order of several km s$^{-1}$ are commonplace for detections in hot Jupiter atmospheres \citep[e.g.][]{Sanchez-Lopez2019}. In those cases, blueshifted signals have been explained theoretically by high-altitude winds propagating from day-to-night side. \citep{Kempton2012}. It is not clear if a similar process could apply for the present temperate sub-Neptune. Other potential explanations could be unaccounted for instrumental effects or model uncertainties. 

It is worth discussing additional structure in the S/N map, especially peaks of comparable strength to the putative CH$_4$ signal. Indeed, the presence of such structure means that more robust observations are needed to definitively confirm or rule out CH$_4$. It is possible spurious peaks and troughs are due to unmitigated stellar or telluric residuals. For example, we performed a close inspection of one spurious peak at $V_{\rm sys} \approx -75$ km s$^{-1}$ and $K_p \approx 100$ km s$^{-1}$. We find significant contribution to this peak from two unmasked orders, between $1.68-1.73 \, \mu$m. These orders contain particularly deep stellar lines which are incompletely removed during SVD detrending. Similarly, strong positive and negative features also appear during the out-of-transit test in Figure~\ref{fig:flip}, which may be attributed to similar causes. Our nominal CH$_4$ signal, however, arises from the full set of detrended orders and is not strongly present in any small subset. Since detrending is performed in the geocentric frame, residuals from deep stellar lines may persist owing to the barycentric correction velocity, leading to spurious cross-correlation peaks which diminish the S/N of the main CH$_4$ signal. We note that spurious structure can also influence the quoted S/N, through the calculation of the baseline standard deviation in \S\ref{subsec:ccf}. We chose a baseline $V_{\rm sys}$ region that appeared clear of gradients in the CCF matrix; however, when extending the baseline significantly (e.g., to $-500$ km s$^{-1}$ to $+500$ km s$^{-1}$), our tentative CH$_4$ signal becomes slightly diminished, with significances of $1.7$ to $2.8\sigma$ at offsets between $-4$ to $-7$ km s$^{-1}$.

Neither the NH$_3$ nor H$_2$O S/N maps show any significant extended features near the expected $K_{\rm p}$ and $V_{\rm sys}$. We performed additional cross-correlation experiments in search of an H$_2$O signal. In particular, we modified our mask in order to admit more spectral orders, including regions with stronger H$_2$O absorption. When including this data in the cross-correlation analysis, we found a feature in excess of $5\sigma$ significance near the expected orbital parameters; however, we believe this to be an artifact from incomplete removal of stellar H$_2$O absorption in the transmission spectra. While absorption by telluric water vapor should be removed during the detrending process (as telluric CH$_4$ was successfully removed), Earth's barycentric motion would have shifted the stellar photospheric H$_2$O lines by $0.34$ km over the course of the night. Stellar lines are also affected, to a lesser extent, by the star's orbital motion around the system barycenter, and by the Rossiter-McLaughlin effect during transit. Indeed, we found that stellar and telluric residuals in the CCF were suppressed only with significant detrending ($n_{\rm SV} \geq 14$). With this level of processing, injected atmospheric H$_2$O signals are significantly degraded. More sophisticated methods may be necessary to make robust HRCCS detections of species shared between the exoplanet, Earth, and host star, for $V_{\rm sys} \approx 0$ km s$^{-1}$.

\subsection{Injection Tests: Cloud-Free Atmosphere}
\label{subsec:injection}

We test our detection sensitivity to these species via injection/recovery tests. The second row of Figure~\ref{fig:maps} shows the results of an identical analysis to the top row, except a model atmospheric transmission spectrum was injected (and cross-correlation was optimized) at $K_{\rm p} = 68$ km s$^{-1}$ and $V_{\rm sys} = 80$ km s$^{-1}$. The model was injected at $1\times$ its nominal strength, after first convolving it to match the spectral resolution of IGRINS. 

For CH$_4$ and NH$_3$ we recover $5.6\sigma$ and $3.7\sigma$ signals, respectively. Therefore, we establish that the observations are indeed sensitive to these two species in the case of a clear atmosphere. This is consistent with similar analyses on low-velocity planets carried out using injections of cloud-free model spectra into this dataset, as reported by Cheverall \& Madhusudhan (in press). Recovery of H$_2$O is achieved at only a $2.9\sigma$ level, which we attribute to its relatively weak features in the unmasked channels. Remarkably, the Doppler shift of the planetary signal throughout transit is only $\Delta v_p = 1.84$ km s$^{-1}$. For hot Jupiter case studies in the HRCCS literature, this value typically reaches tens of km s$^{-1}$ \citep[e.g.][]{Brogi2018}. Nevertheless, our injected CH$_4$ and NH$_3$ signals remain preserved through the detrending process. 

An argument may be made that $V_{\rm sys} \sim 0$ presents further challenges to atmospheric detection, since atmospheric spectral features will more closely align with the cores of telluric lines. The atmospheric signal may therefore have lower S/N, or be more degraded by detrending. We therefore repeated the above injection test using the orbital parameters $K_{\rm p} = -68$ km s$^{-1}$ and $V_{\rm sys} = 0$ km s$^{-1}$. This injected signal may more closely mimic the statistical properties of the true atmospheric transmission spectrum of this target. CH$_4$ and NH$_3$ are still recovered, albeit at the lower significances of $2.6\sigma$ and $1.6\sigma$, respectively. H$_2$O is recovered at the $1.2\sigma$ level.

We therefore conclude that the combination of a low-significance inference of CH$_4$ and no indication of NH$_3$ in \S\ref{subsec:molecular_inferences} are consistent with more CH$_4$ absorption relative to NH$_3$ in the atmosphere. Alternatively, if the CH$_4$ signal is spurious, then both species may be depleted, or the atmosphere may be cloudy (discussed in \S\ref{subsec:injection_cloudy}). The latter scenario could also point toward an atmosphere which is not H$_2$-dominated, or the species being photochemically destroyed. 

\subsection{Injection Tests: Cloudy Atmosphere}
\label{subsec:injection_cloudy}

We additionally consider the possibility of high-altitude clouds in the atmosphere of the planet which could be obscuring the molecular signatures. In particular, we investigate whether the low-significance inference of CH$_4$ in the data can be explained by the presence of high-altitude clouds. To test this scenario, we perform injection-recovery tests using CH$_4$-only models with a gray cloud deck with different cloud-top pressures ($P_{\rm c}$), spanning 10$^{-2}$ - 10$^{-6}$ bar, with lower pressures implying higher in the atmosphere. The CH$_4$ abundance was set at 1\%. We inject the models at $V_{\rm sys}$ = $80$ km s$^{-1}$ and $K_{\rm p} = 68$ km s$^{-1}$, and, as in the previous sections, attempt to recover the signal using $\Delta$CCF optimization for each model individually. The results are shown in Figure~\ref{fig:maps_cloudy}. 

We find that the detection significance of the cloudy model decreases with the cloud-top pressure, which is consistent with expectations. At the lower end with $P_{\rm c} = 10^{-2}$ bar, or 10 mbar, the detection significance is lowered to 4.7$\sigma$, i.e., nearly 1$\sigma$ lower than the equivalent cloud-free CH$_4$-only model. Importantly, for $P_{\rm c}$ below 10$^{-4}$ bar, the detection significance is under 3$\sigma$.  As such, our 2$\sigma$ significance with the CH$_4$-only model is arguably consistent with the presence of a high-altitude cloud deck at $P_{\rm c}$ below 0.1 mbar. We also conduct similar injection tests for NH$_3$ and find that the detection significance reduces to the noise level for a cloud deck below 0.1 mbar, also consistent with our cross-correlation results. 

\section{Summary and Discussion}
\label{sec:disc}

We report a ground-based reconnaissance program of a temperate sub-Neptune TOI-732~c using high-resolution transit spectroscopy in the near-infrared. The planet orbits a bright (J=9) nearby M dwarf making it a promising target for atmospheric characterisation. In the present study, we investigate the feasibility of characterising its atmosphere using time-series observations with the IGRINS instrument on Gemini-South, conducted over one primary transit of the planet in front of its host star. Our successful observations, which were conducted under near-ideal conditions, resulted in excellent data quality and provide important insights into the viability of characterising temperate sub-Neptunes using ground-based 8-m class telescopes. 

We use the HRCCS technique to search for prominent molecular features in the planetary atmosphere. The bulk density of the planet is consistent with the presence of an H$_2$-rich atmosphere or a steam atmosphere \citep[e.g.][]{Madhusudhan2021,Bonfanti2023}. Therefore, atmospheric observations are essential to break this degeneracy. We search for the prominent CNO molecules expected in temperate H$_2$-rich atmospheres, namely, H$_2$O, CH$_4$ and NH$_3$.

We report initial insights into the atmospheric composition of TOI-732~c. We find nominal evidence, at $2.2\sigma$, for the presence of CH$_4$ in a H$_2$-rich atmosphere and no evidence for NH$_3$. We conduct injection tests to establish the robustness of these results, and find that both CH$_4$ and NH$_3$ would be detectable at high significance, if present in expected quantities in a cloud-free H$_2$-rich atmosphere. The low detection significance for CH$_4$ and non-detection of NH$_3$ may indicate that the atmosphere may have an inherently high CH$_4$/NH$_3$ ratio in the observable photosphere in transit geometry. Alternatively, it may indicate the presence of a high-altitude cloud deck in the atmosphere partially masking the spectral features, consistent with our injection tests with cloudy atmospheric models. Finally, we do not detect H$_2$O, as our search is hindered by contamination from telluric water vapor, especially given the low $V_{\rm sys}$, as well as possible water in parts of the stellar photosphere.

Beyond the above findings, our results demonstrate the feasibility of chemical detections in temperate sub-Neptunes using high-resolution spectroscopy from ground. Such planets are usually associated with two challenges. Firstly, the low temperature and small size of such planets lead to small atmospheric signatures, which are hard to observe in general, from space or on ground. Additionally, the presence of high-altitude clouds can further inhibit the spectral signature. Secondly, the low temperature is associate with a long orbital period, implying a small radial velocity shift during transit. This is unfavourable to high-resolution cross-correlation spectroscopy from ground which typically relies on high velocity dispersion.

The present results demonstrate that both of the above challenges can be surmounted using a high-throughput spectrograph on an 8-m class telescope. The first challenge is addressed by the high S/N of the observations, thanks to the instrument, near-ideal observing conditions and a bright M dwarf host, and a H$_2$-rich sub-Neptune atmosphere. The second challenge is addressed by the fact that even for a low radial velocity the planetary signal is present during the transit but not out of transit, whereas the stellar and telluric signals are present is present in all frames. As such, with adequate masking and detrending of the static signals, the unique planetary features during transit can be detected reliably. A more general exploration of this approach is discussed in Cheverall \& Madhusudhan (in press). Finally, we note that our findings are based on a single transit observation, and similar observations over multiple transits and/or those with other facilities (e.g. JWST) could verify the present inferences and improve the detection significance. 

Our findings present important implications for JWST observations of the planet, considering the recent detections of carbon-bearing molecules in a comparable Hycean candidate K2-18~b \citep{Madhusudhan2023b}. In particular, our tentative evidence for CH$_4$ and non-detection of NH$_3$ can be confirmed with upcoming JWST observations (GO Program 3557, PI: N. Madhusudhan). A plausible scenario for TOI-732~c is that CH$_4$ is present in the atmosphere, while NH$_3$ is significantly depleted relative to equilibrium chemistry expectations. Such a scenario has been found to be the case for the temperate sub-Neptune K2-18~b, where JWST observations revealed significant quantities of CH$_4$ in its atmosphere and a paucity of NH$_3$ \citep{Madhusudhan2023b}. In the case of K2-18~b, these constraints, along with the presence of CO$_2$, were interpreted as consistent with the presence of an ocean underneath the H$_2$-rich atmosphere suggesting it as a possible Hycean world. 

We also find that our chemical inferences for TOI-732~c are consistent with the presence of high-altitude clouds muting spectral features. This means that JWST observations using the NIRISS spectrograph, whose wavelength coverage encompasses the IGRINS band, may be expected to show similarly muted spectral features. Such muting may therefore hinder the detectability of CH$_4$ and NH$_3$ if relying on NIRISS data alone. However, TOI-732~c is also set to be observed with NIRSpec G395H and MIRI, which together span a $\sim$3-10~$\mu$m range. As such, there may be windows in the cloud aerosol opacity over this large wavelength range which can enable robust atmospheric characterisation, as indicated by prior studies using simulated \citep[e.g.][]{Wakeford2015, Pinhas2017} and actual JWST observations \citep[e.g.][]{Constantinou2023}. Moreover, the variation of the spectral muting with wavelength can be used by atmospheric retrievals to additionally constrain the properties of the aerosols comprising the clouds, such as their composition, spatial distribution and modal particle size.

Altogether, our reconnaissance of TOI-732~c represents a successful case study and important milestone. It sets precedent for future investigations of sub-Neptunes orbiting nearby M dwarfs, using 8m class, high-resolution ground-based facilities. We have demonstrated the viability of GEMINI-S/IGRINS to probe sub-Neptune atmospheres even in the presence of high-altitude clouds, and to confirm such planets as prime targets for further atmospheric characterisation with JWST. 

\begin{acknowledgments}

This work is supported by the UK Research and Innovation (UKRI) Frontier Grant (EP/X025179/1, PI: N. Madhusudhan). The observations used in this work were conducted as part of Gemini-S program GS-2021A-Q-201 (PI: D. Valencia). N.M. and S.C. acknowledge support from the UKRI Frontier grant. S.H.C.C. acknowledges support from the Junior Research Fellowship at Peterhouse, University of Cambridge. D.V. acknowledges support from the Natural Sciences and Engineering Research Council of Canada (grant RGPIN-2021-02706). J. M. V. acknowledges support from a Royal Society - Science Foundation Ireland University Research Fellowship (URF$\backslash$1$\backslash$221932). T.M. acknowledges financial support from the Spanish Ministry of Science and Innovation (MICINN) through the Spanish State Research Agency, under the Severo Ochoa Program 2020-2023 (CEX2019-000920-S). N.M. and C.C. acknowledge support from the MERAC Foundation, Switzerland, and the UK Science and Technology Facilities Council (STFC) Center for Doctoral Training (CDT) in Data Intensive Science at the University of Cambridge, towards the doctoral studies of C.C. The authors thank the NOIRLabs support staff and the Immersion Grating Infrared Spectrometer (IGRINS) instrument team for helping with the implementation of these observations and the data reduction. NM thanks Gregory Mace of the IGRINS team for the very helpful discussions. We thank the anonymous referee for their valuable comments which helped improve the manuscript. 

This program, GS-2021A-Q-201, is based on observations obtained at the international Gemini Observatory, a program of NSF’s NOIRLab, which is managed by the Association of Universities for Research in Astronomy (AURA) under a cooperative agreement with the National Science Foundation on behalf of the Gemini Observatory partnership: the National Science Foundation (United States), National Research Council (Canada), Agencia Nacional de Investigación y Desarrollo (Chile), Ministerio de Ciencia, Tecnología e Innovación (Argentina), Ministério da Ciência, Tecnologia, Inovações e Comunicações (Brazil), and Korea Astronomy and Space Science Institute (KASI, Republic of Korea). This work used the IGRINS instrument that was developed under a collaboration between the University of Texas at Austin and KASI with the financial support of the Mount Cuba Astronomical Foundation, of the US National Science Foundation under grants AST-1229522 and AST-1702267, of the McDonald Observatory of the University of Texas at Austin, of the Korean GMT Project of KASI, and Gemini Observatory. 

This research has made use of the NASA Exoplanet Archive, which is operated by the California Institute of Technology, under contract with the National Aeronautics and Space Administration under the Exoplanet Exploration Program. This research has made use of the NASA Astrophysics Data System and the Python packages \texttt{NUMPY}, \texttt{SCIPY}, and \texttt{MATPLOTLIB}.

\end{acknowledgments}

{\it Author Contributions:} N.M. conceived, planned and led the project with contributions from all authors as follows. N.M., D.V., S.C., J.M.V. and T.M. contributed to the successful observing proposal. N.M., T.M. and J.M.V. planned the observations. S.H.C.C. and N.M. conducted the data analysis. S.C. and N.M. conducted the atmospheric modelling. S.H.C.C, S.C. and N.M. led the writing of the manuscript with comments and edits from all authors.

\bibliography{ms.bib,references.bib}
\bibliographystyle{aasjournal}

\begin{appendix}

\section{Considerations for Night 1 Data} \label{sec:n1}

\subsection{Data Quality}

\begin{figure*}[t]
\noindent
\includegraphics[angle=0,width=\textwidth]{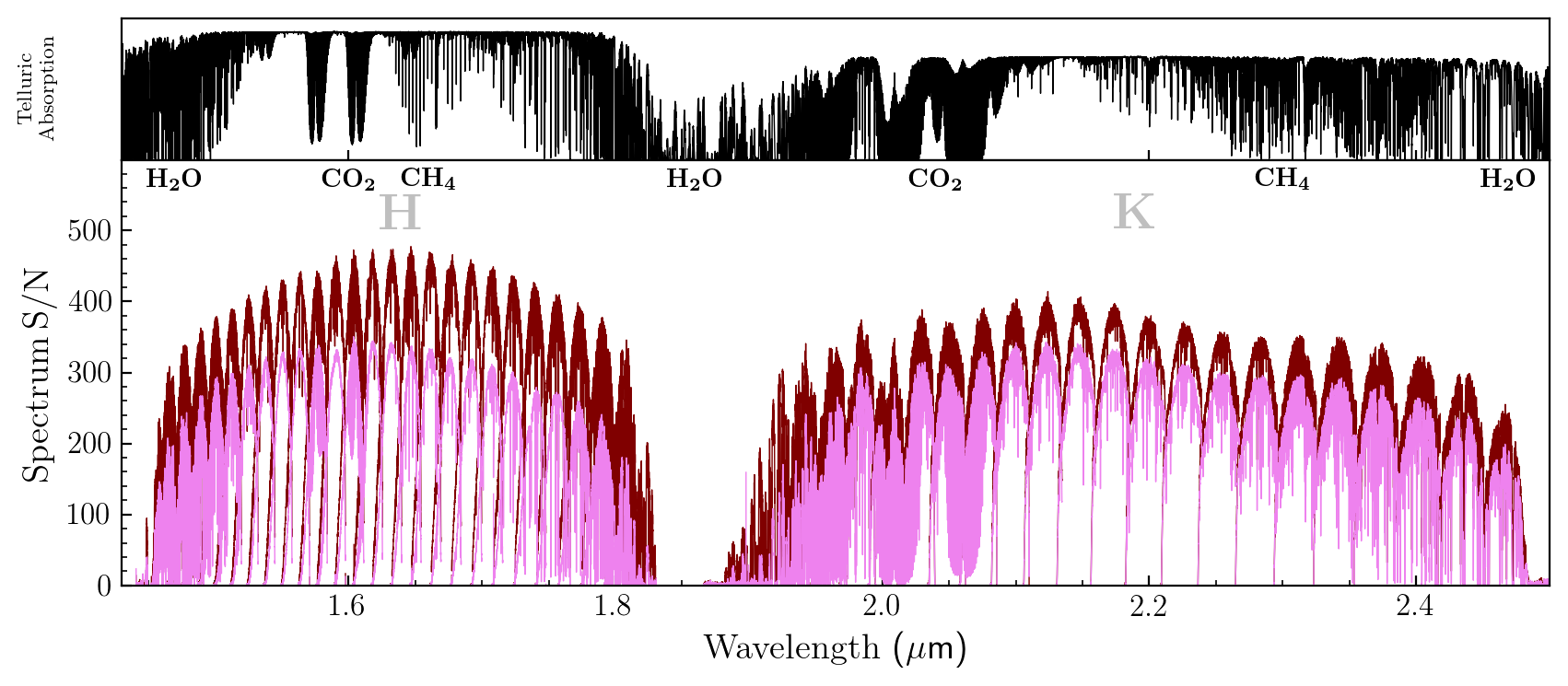}
\caption{Top: Fiducial telluric spectrum, as in Figure~\ref{fig:snr}. Bottom: S/N of first exposure obtained on Night 1 (violet), which was excluded from our analysis. The Night 1 spectra have lower S/N compared to Night 2 (maroon) and do not have adequate out-of-transit frames at low airmass.
}
\label{fig:snrn1}
\end{figure*}

\begin{figure*}[t]
\noindent
\centering
\includegraphics[angle=0,width=0.5\textwidth]{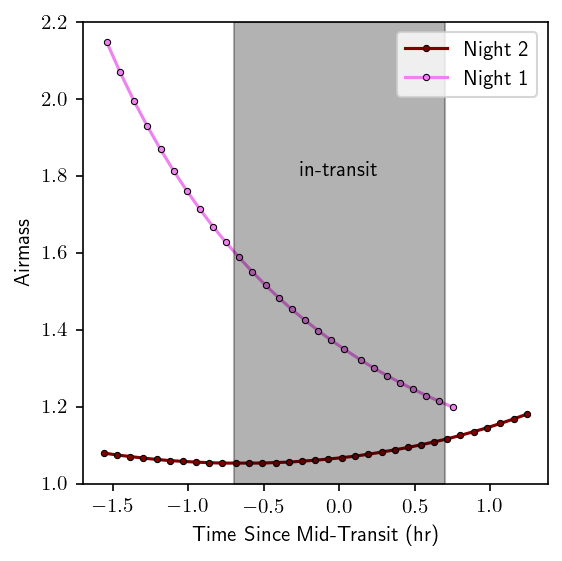}
\caption{Airmass as function of time for Night 1 (violet) and Night 2 (maroon). Exposures during Night 1 took place at significantly higher airmass, which is one of the reasons for the degradation in S/N (Figure~\ref{fig:snrn1}). Moreover, Night 1 yielded only one post-transit exposure.}
\label{fig:amc}
\end{figure*}

For completion, we present spectra attained on Night 1; although, these spectra were excluded from the primary analysis of this manuscript. In total, 27 4-minute (A/B pair) exposures were attained. The exposures capture the full transit of TOI-732~c, but only one exposure occurred during the post-transit baseline. The data are of generally lower quality compared to Night 2 as the typical S/N per pixel is $\sim 200$ (Figure~\ref{fig:snrn1}). This is in part due to the substantially higher airmass of the observations (Figure~\ref{fig:amc}), ranging $z=2.15 \rightarrow 1.20$, which also induces severe H$_2$O telluric contamination. We tested for the presence of CH$_4$ with our cross-correlation pipeline, but did not identify a statistically significant signal near the expected $V_{\rm sys}$ or $K_{p}$ of the system. Furthermore, when we injected our model CH$_4$ transmission spectrum, the cross-correlation signal was consistently recovered at low confidence ($\lesssim 1.5\sigma$), for injections at both $V_{\rm sys} = 80$ km s$^{-1}$ and $V_{\rm sys} = 0$ km s$^{-1}$. These poor statistical properties, in conjunction with Night 1's severe telluric contamination, led us to exclude it from our nominal analysis.

\section{Additional Robustness Tests} \label{sec:oot}

\subsection{Effect of Out-Of-Transit Frames}

\begin{figure*}[t]
\noindent
\includegraphics[angle=0,width=\textwidth]{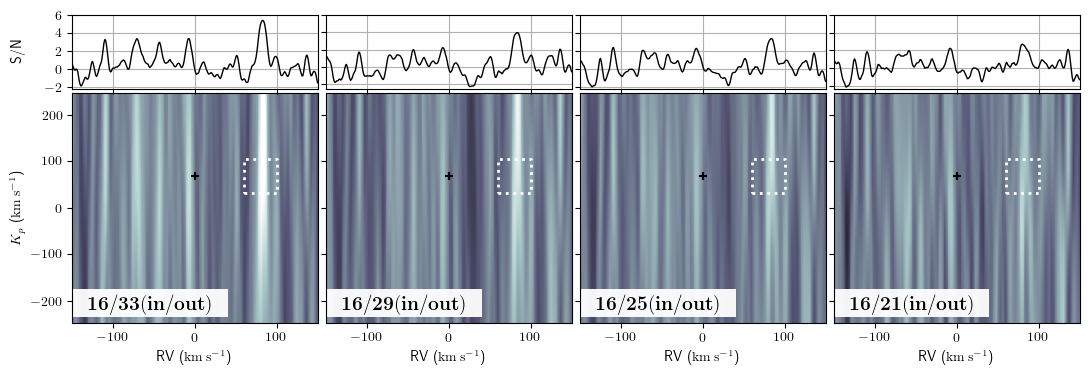}
\caption{Recovery of a synthetic signal injected at $V_{\rm sys}=80$ km s$^{-1}$ and $K_{p}=68$ km s$^{-1}$ (denoted by the dotted box), based on our detrending and cross-correlation procedure described in \S\ref{sec:obs}. The leftmost panel shows the nominal recovery using all frames. In the subsequent panels we have reduced the number of out-of-transit frames, which has the effect of gradually weakening the significance of the recovered signal (from 5.5$\sigma$ nominally, to 3.2$\sigma$ when only 5 of the available 17 out-of-transit frames are included). For all panels, the color-scale is normalized between $-5\sigma$ and $+5\sigma$, and the top panel shows the 1D slice at $K_{p}=68$ km s$^{-1}$.
}
\label{fig:nouttrend}
\end{figure*}

The necessity of out-of-transit baseline exposures for isolating the atmospheric signal is established. As discussed in \S\ref{subsec:detrending}, SVD can flexibly identify and remove telluric and stellar features, but runs a risk of degrading the atmospheric signal \citep{Birkby2013, deKok2013} especially with a higher number of removed singular vectors. This risk is present in our analysis of TOI-732~c, whose low semi-amplitude induces only a moderate Doppler shift across in-transit frames. Although, our significant out-of-transit baseline helps mitigate this effect. Approximately half of our exposures are out-of-transit, which helps prevent the atmospheric signal from being captured in the leading singular components. The effect of fewer out-of-transit exposures is shown in Figure~\ref{fig:nouttrend}, via the recovery of a synthetic atmospheric signal injected into the data at $V_{\rm sys}=80$ km s$^{-1}$ and $K_{p}=68$ km s$^{-1}$ prior to the analysis. Out-of-transit exposures are gradually cropped from either end of the time-series and, as a result, the significance of the recovered signal is substantially reduced.

\subsection{Bootstrapped Cross-Correlation} \label{subsec:bootstrapped}

A bootstrap analysis is conducted to further test the robustness of our tentative CH$_4$ signal. Spectra are cleaned, continuum-normalized, and assigned orbital phases and in-transit/out-of-transit designations, per \S\ref{sec:obs}. In one bootstrap trial, we resample (with replacement) the spectral timeseries. Specifically, the fluxes in each wavelength channel of one spectrum are replaced with the fluxes in the corresponding channels of a different, randomly selected spectrum, and this process is repeated for all 33 spectra in the timeseries. In each case, the replacement spectrum may be selected from either the original in-transit or out-of-transit sample. Detrending with $\Delta$CCF optimization is then performed, followed by cross-correlation. The S/N values across pairs of $K_{p}$, $V_{\rm sys}$ are recorded. We repeat for 1000 bootstrap trials.

\begin{figure*}[t]
\noindent
\centering
\includegraphics[angle=0,width=0.5\textwidth]{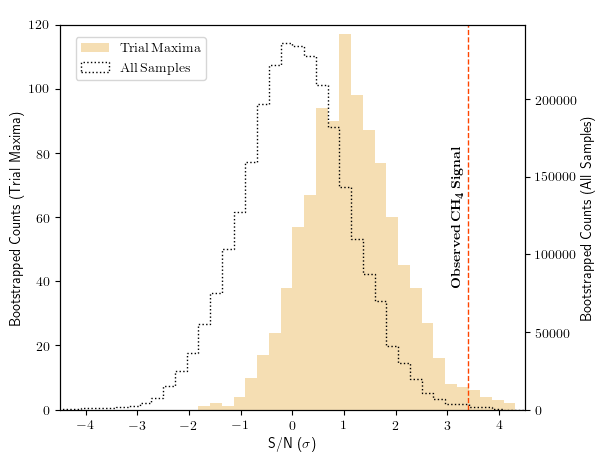}
\caption{Results of the bootstrapped cross-correlation analysis. This figure plots the distribution of S/N values at $K_{p}$, $V_{\rm sys}$ pairs near the tentative CH$_4$ signal, for all 1000 bootstrap trials (where, in each trial, spectra were randomly resampled with replacement at the start of the cross-correlation analysis). The distribution of all S/N values (``All Samples") is shown by the dotted curve, whereas the filled tan histogram (``Trial Maxima") shows the distribution of maxima from each trial. The peak S/N of the CH$_4$ signal is marked by a vertical dashed line.}
\label{fig:bootstrap}
\end{figure*}

For each bootstrap trial, we isolate S/N values lying in the range $-10 \, {\rm km \, s^{-1}} \leq V_{\rm sys} \leq +1 \, {\rm km \, s^{-1}}$, and $+30 \, {\rm km \, s^{-1}} \leq K_{p} \leq +250 \, {\rm km \, s^{-1}}$. This restriction acts as an approximate prior on velocities. That is, if a statistically significant signal was detected in this range, it may be plausibly atmospheric in origin. The range roughly accounts for systematic and model uncertainties, and the statistical properties of the data. Next, we benchmark our CH$_4$ signal to the distribution of bootstrapped S/N values. In Figure~\ref{fig:bootstrap}, we plot two distributions: the dotted profile (``All Samples Distribution") corresponds to all 2,652,000 S/N values resulting from all bootstrap trials, and the subsequent velocity range restriction. The tan histogram (``Trial Maxima Distribution") corresponds to the maximum of S/N values in each bootstrap trial (here, there are 1000 total samples, one for each bootstrap trial). With the observed CH$_4$ signal peaking at S/N = $3.4\sigma$, $0.15\%$ of samples in the ``All Samples Distribution" lie at greater S/N. Under a normal distribution, this position in the tail implies a $3.0\sigma$ outlier. Separately, $1.5\%$ of samples in the ``Trial Maxima Distribution" lie at greater S/N than our tentative CH$_4$ signal, which corresponds to a $2.2\sigma$ outlier. Therefore, the 3.4 S/N of the CH$_4$ inference is well calibrated to the statistical properties of the data. Furthermore, the peak remains an outlier across bootstrap trials, suggesting there are no trends or artifacts in the data that would consistently generate this signal through resampling spectra.

\setlength{\tabcolsep}{3pt}
\begin{table}
\centering
\caption{Number of singular vectors ($n_{\rm SV}$) removed for each spectral order, determined by $\Delta$CCF opimization for our CH$_4$ atmospheric model. Hyphens (``-") indicate the order was fully masked and not used in cross-correlation.}
\begin{tabular}{||c | c c c c c c c c c c c c c c c c c c c c c c c c c c c||} 
 \hline
Order & 1 & 2 & 3 & 4 & 5 & 6 & 7 & 8 & 9 & 10 & 11 & 12 & 13 & 14 & 15 & 16 & 17 & 18 & 19 & 20 & 21 & 22 & 23 & 24 & 25 & 26 & 27 \\
$n_{\rm SV}$ & - & - & - & - & - & - & 5 & 2 & 1 & 1 & 1 & 9 & 1 & 1 & 2 & 2 & 5 & 1 & 2 & 2 & 1 & 1 & 1 & 1 & - & - & - \\
 \hline\hline
 Order & 28 & 29 & 30 & 31 & 32 & 33 & 34 & 35 & 36 & 37 & 38 & 39 & 40 & 41 & 42 & 43 & 44 & 45 & 46 & 47 & 48 & 49 & 50 & 51 & 52 & 53 & 54 \\
$n_{\rm SV}$ & - & - & - & - & - & - & - & - & - & - & - & - & 2 & 3 & 1 & 3 & 5 & 3 & 9 & 3 & 10 & - & - & - & - & - & - \\
 \hline
\end{tabular}
\label{tab:svd}
\end{table}

\end{appendix}

\end{document}